\documentclass{ws-ijmpb}

\def\ZzZ{{\hbox{\tenrm Z\kern-.31em{Z}}}}
\def\CcC{{\hbox{\tenrm C\kern-.45em{\vrule height.67em width0.08em depth-
.04em \hskip.45em }}}}

\def\mapbelow#1{\smash{\mathop{\longrightarrow}\limits_{#1}}}

\newcommand{\lab}{\label}

\newcommand{\bc}{\begin{center}}
\newcommand{\ec}{\end{center}}
\newcommand{\be}{\begin{equation}}
\newcommand{\ee}{\end{equation}}
\newcommand{\bea}{\begin{eqnarray}}
\newcommand{\eea}{\end{eqnarray}}
\newcommand{\bs}{\begin{subequations}}
\newcommand{\es}{\end{subequations}}
\newcommand{\beq}{\begin{eqalignno}}
\newcommand{\eeq}{\end{eqalignno}}

\def\lab{\label}

\def\lf{\left}

\def\ri{\right}

\def\lab{\label}

\begin{document}

\markboth{Giuseppe Vitiello}
{Classical chaotic trajectories in
QFT}

%
\catchline{}{}{}{}{}
%

\title{Classical chaotic trajectories in quantum field theory
}

\author{Giuseppe Vitiello
}

\address{Dipartimento di Fisica ``E.R.Caianiello", Universit\`a di Salerno,
84100 Salerno, Italy\\ 
INFN, Gruppo Collegato di Salerno and INFM, Sezione di Salerno\\ 
vitiello@sa.infn.it}

%
%

\maketitle


\begin{abstract}
Trajectories in the space of the unitarily inequivalent
representations of the canonical commutation relations are shown
to be classical trajectories. Under convenient conditions, they
may exhibit properties typical of chaotic behavior in classical
nonlinear dynamics. Quantum noise in fluctuating random force in
the system--environment coupling and system--environment
entanglement is also discussed.
\end{abstract}


\section{Introduction
}


In recent years the structure of the state space in quantum field
theory (QFT) has been shown\cite{Celeghini:1998sy} to be
intimately related to the one of the deformed Hopf
algebra\cite{25.,Celeghini:1991km}. This fact has been exploited
to clarify some characteristic features of thermal field theories
and it has been shown that the couple of ``thermal" conjugate
variables $\theta$ and $p_{\theta} \equiv
-i\frac{\partial}{\partial \theta }$, with $\theta$ related to the
$q$--deformation parameter, may describe trajectories in the space
$\cal H$ of the representations, i.e. the space whose ``points"
are the unitarily inequivalent representations (uir) of the
canonical commutation relations (ccr)\cite{Celeghini:1998sy}. In
the present paper I will show that there is a symplectic structure
associated to the ``thermal degrees of freedom" $\theta$ and that
the trajectories in the $\cal H$ space may exhibit some properties
typical of chaotic trajectories in classical nonlinear dynamics.

In Section 2 I will shortly summarize the deformed Hopf algebra
structure underlying QFT. In Section 3 I will comment on the
quantum noise nature of the doubling of degrees of freedom
implicit in the coproduct mapping of the Hopf algebra and on the
entanglement between the system degrees of freedom and the doubled
ones. In Section 4 I will show that trajectories in the $\cal H$
space are classical trajectories, which, under convenient
conditions, may satisfy the criteria for chaoticity prescribed by
nonlinear dynamics. For shortness I will not discuss possible
applications of the formalism here presented.


I am pleased to dedicate to Francesco Guerra these notes on QFT:
in some sense my interest in the issues here discussed is rooted
in my thesis work towards the degree in Physics supervised by him
at the Naples University ``some" time ago.

\section{Hopf algebra and the doubling of the degrees of freedom}

In the following I will consider the case of bosons. The
conclusions, however, apply also to the fermion case.

The additivity of so-called primitive operators, such as energy,
momentum, angular momentum, necessarily implies the use of the
Lie-Hopf algebra in quantum theories. The prescription for
operating on two modes is indeed provided by the coproduct
operation, a key ingredient of Hopf algebras. The addition of,
e.g., the angular momentum $J^{\alpha}$, ${\alpha} = 1,2,3$, of
two particles, is given by the coproduct $\Delta J^{\alpha} =
J^{\alpha} \otimes {\bf 1} + {\bf 1} \otimes J^{\alpha} \equiv
J^{\alpha}_1 + J^{\alpha}_2$. The coproduct  is a homomorphism
which duplicates the algebra, ${\Delta}: {\cal A}\to {\cal
A}\otimes {\cal A}$, i.e. $\Delta {\cal O} = {\cal O} \otimes {\bf
1} + {\bf 1} \otimes {\cal O} \equiv {\cal O}_1 + {\cal O}_2$,
with ${\cal O} \in {\cal A}$.

A remarkable result is that the infinitely many uir of the ccr,
whose existence characterizes QFT, are classified by use of the
deformed Hopf algebra. Quantum deformations of Hopf algebra have
thus a deeply non-trivial physical meaning in QFT.  One can indeed
show\cite{Celeghini:1998sy,Iorio:1994jk} that the Bogolubov
transformations
\bea A(\theta) \equiv \frac{1}{\sqrt{2}} \left ( {\alpha}(\theta )
+ {\beta}(\theta )\right ) &=& A ~{\rm cosh} ~\theta -
{B}^{\dagger}
~{\rm sinh} ~\theta ~~, \lab{p321a} \\
B(\theta) \equiv \frac{1}{\sqrt{2}} \left ( {\alpha}(\theta ) -
{\beta}(\theta ) \right ) &=& B ~{\rm cosh} ~\theta - A^{\dagger}
~{\rm sinh} ~\theta ~, \lab{p321} \eea
which relate different (i.e. unitary inequivalent)
representations\cite{BR}, are directly obtained by use of the
deformed copodruct operation:
\be \Delta a_{q} =  a_1 q^{1/2} + q^{-1/2} a_2 ~,~~~ ~\Delta
a_{q}^{\dagger} = a_1^{\dagger} q^{1/2}  +q^{-1/2}  a_2^{\dagger}
~.\lab{p213} \ee
Note that $[a_i , a_j ] = [a_i , a_{j}^{\dagger} ] = 0 , ~i,j =
1,2,~ i \neq j $. In Eqs. (\ref{p321a}) and (\ref{p321})
${\alpha}(\theta)$ and ${\beta}(\theta )$ are convenient linear
combinations\cite{Celeghini:1998sy} of the coproduct operators
(\ref{p213}) with $q = e^{2\theta}$. Note that $[ A(\theta) ,
A^{\dagger}(\theta) ] = 1 ~, ~~[ B(\theta) , B^{\dagger}(\theta) ]
= 1 $ and all other commutators equal to zero. $A(\theta)$ and
$B(\theta)$ also commute. The momentum suffix $\kappa$ is omitted
for simplicity.

Since QFT is characterized by the existence of uir of the ccr, the
intrinsic algebraic structure of QFT is thus the one of the
deformed Hopf algebra.

The generator of (\ref{p321a}) and (\ref{p321}) is ${\cal G}
\equiv -i(A^{\dagger}B^{\dagger} - AB)$:
\be - i{\delta \over {\delta \theta}} A(\theta) = [{\cal G},
A(\theta)] ~,~~~ - i{\delta \over {\delta \theta}} B(\theta) =
[{\cal G}, B(\theta)] ~, \lab{p326}\ee
and h.c.. Thus $\displaystyle{p_{\theta} \equiv -i{\delta \over
{\delta \theta}}}$ can be regarded\cite{Celeghini:1998sy} as the
momentum operator ``conjugate" to the ``degree of freedom"
$\theta$. For an assigned fixed value $\bar{\theta}$, it is
\be \exp(i{\bar{\theta}} p_{\theta}) ~A(\theta) =
\exp(i{\bar{\theta}}{\cal G}) ~A(\theta)~
\exp(-i{\bar{\theta}}{\cal G}) = A( \theta + {\bar {\theta}} ) ~,
\lab{p327}\ee
and similarly for $B(\theta)$.

It is interesting to consider the case of time--dependent
deformation parameter. In such a case, the Heisenberg equation for
$A(t,{\theta} (t))$ is
$$
-i{\dot A}(t,{\theta}(t)) = -i{\delta \over {\delta t}}
A(t,{\theta}(t)) -i{{\delta \theta} \over {\delta t}}~ {\delta
\over {\delta \theta}}A(t, {\theta}(t))=
$$
%
\be \left [ H , A(t,{\theta}(t)) \right ] + {{\delta \theta} \over
{\delta t}}~ [{\cal G}, A(t,{\theta}(t)) ] = \left [ H + Q ,
~A(t,{\theta}(t)) \right ] ~, \lab{42} \ee
and $\displaystyle{Q \equiv {{\delta \theta} \over {\delta t}}
{\cal G}}$ plays the role of the heat--term in dissipative
systems. $H$ is the Hamiltonian responsible for the time variation
in the explicit time dependence of $A(t,{\theta}(t))$.  $H + Q$
can be therefore identified with the free
energy~\cite{Celeghini:1992yv}: variations in time of the
deformation parameter involve dissipation.

Let $|0\rangle \equiv |0\rangle \otimes |0\rangle$ denote the
vacuum annihilated by $A$ and $B$, $A|0\rangle = 0 = B|0\rangle $.
By introducing the suffix $\kappa$, at finite volume $V$ one
obtains
\be\label{19}
 |0 (\theta) \rangle \, = \exp(i \sum_{\kappa}\theta_{\kappa}
 {\cal G_{\kappa}}) |0 \rangle \,
  =\prod_k\;\frac{1}{\cosh\theta_{k}}\,\exp\left({\tanh\theta_{k}
 A_k^{\dagger} {B}_{k}^{\dagger}}\right)\, |0\rangle \,.
\ee
Here $\theta$ denotes the set $\{\theta_{\kappa} = \frac{1}{2}\ln
q_{\kappa}, \forall \kappa \}$ and $\langle
0(\theta)|0(\theta)\rangle = 1$.

The vacuum $|0 (\theta) \rangle$ is an $SU(1,1)$ generalized
coherent state\cite{perelomov} (the group structure actually is
$\bigotimes_{\kappa} SU(1,1)_{\kappa}$): coherence and the vacuum
structure in QFT are thus intrinsically related to the deformed
Hopf algebra. In the following ${\cal H}_\theta$ will denote the
Hilbert space with vacuum $|0 (\theta) \rangle$: ${\cal H}_\theta
\equiv \{ |0 (\theta) \rangle \}$.

In the infinite volume limit, the number of degrees of freedom
becomes uncountable infinite, hence one
obtains\cite{Celeghini:1992yv,11.} $ \langle
0(\theta)|0(\theta^{\prime})\rangle\to 0$ ~ as $V\to\infty, \quad
\forall
  \theta, \theta^{\prime}, ~ \theta\ne \theta^{\prime}$, which means
that the Hilbert spaces ${\cal H_{\theta }}$ and ${\cal H}_{\theta
'}$ become unitarily inequivalent. In this limit, the ``points" of
the space ${\cal H} \equiv \{ {\cal H}_\theta, ~ \forall \theta
\}$  of the infinitely many uir of the ccr are labelled by the
deformation parameter $\theta$
${}$\cite{Celeghini:1998sy,Iorio:1994jk,Celeghini:1992yv}.

\section{Quantum noise and entanglement}

The  doubling of the degrees of freedom which, as we have seen, is
built in in the algebraic structure of QFT, is rich of physical
meanings (e.g. in thermal field theory the doubled degree of
freedom describes the heat
bath\cite{Celeghini:1998sy,Celeghini:1992yv}; or, near a black
hole, such a doubling describes the modes on the two sides of the
horizon \cite{Martellini:1978sm} ).

The doubling of the degrees of freedom can also be formally
understood by considering the standard expression for the Wigner
function \cite{FeynmanStat},
\be\lab{W} W(p,x,t) = \frac{1}{2\pi \hbar}\int {\psi^* \lf(x -
\frac{1}{2}y,t\ri)\psi \lf(x + \frac{1}{2}y,t\ri)
e^{\lf(-i\frac{py}{\hbar}\ri)}dy} ~. \ee
By using $x_{\pm}=x\pm \frac{1}{2}y$, the associated density
matrix function is
\be\lab{8} W(x,y,t)=\langle x_{+}|\rho (t)|x_{-}\rangle = \psi^*
(x_{-},t)\psi (x_{+},t)~. \ee

The density matrix and the Wigner function formalism thus {\it
requires} the introduction of a ``doubled" set of coordinates,
$(x_{\pm}, p_{\pm})$ (or $(x,p_{x})$ and $(y,p_{y})$). Such a
doubling can be shown\cite{Srivastava:1995yf} to
coincide with the one discussed in Sec. 2.

The doubled degrees of freedom play the role of the thermal bath
(the environment) degrees of freedom\cite{11.}. In the frame of
the formalism by Schwinger\cite{schwinger} and by Feynman and
Vernon \cite{Feynman} for the  quantum Brownian motion it is
possible to show\cite{Srivastava:1995yf} that the role of the
``doubled" $y$ coordinate is absolutely crucial in the quantum
regime, since there it accounts for the quantum noise in the
fluctuating random force in the system-environment coupling: in
the limit of $y \rightarrow 0$ (i.e. for $x_{+} = x_{-}$) quantum
effects are lost and the classical limit is obtained.

I now remark that the state $|0(\theta)\rangle$ in Eq. (\ref{19})
is an entangled state. The entanglement between the modes $A$ and
$B$ is indeed manifest by writing
\bea\label{ent}
 |0 (\theta) \rangle \,
= \left ( \prod_k\;\frac{1}{\cosh\theta_{k}} \right)\,\left ( |0
\rangle \otimes |{0} \rangle + \sum_k \;\tanh\theta_{k}
  \left( | A_k  \rangle \otimes |{B}_{k} \rangle
  \right)  + \dots \right)~,
\eea
which clearly  cannot be factorized into the product of two
single-mode states. By defining $ |\,{\cal I}\rangle \, \equiv
\exp {\left( \sum_{\kappa} A_{\kappa}^{\dagger}
{B}_{\kappa}^{\dagger} \right)} |0\rangle $, the state $|0
(\theta) \rangle $ may be also written as:
\be |0 (\theta) \rangle \, = \exp{\left ( - {1\over{2}} S_{A}
\right )} |\,{\cal I}\rangle  \, = \exp{\left ( - {1\over{2}}
S_{B} \right )} |\,{\cal I}\rangle  ~, \ee
\be S_{A} \equiv - \sum_{\kappa} \Bigl \{ A_{\kappa}^{\dagger}
A_{\kappa} \ln \sinh^{2} {\theta}_{\kappa} - A_{\kappa}
A_{\kappa}^{\dagger} \ln \cosh^{2} {\theta}_{\kappa} \Bigr \} ~.
\ee

$S_{B}$ is given by a similar expression  with ${B}_{\kappa}$ and
${B}_{\kappa}^{\dagger}$ replacing $A_{\kappa}$ and
$A_{\kappa}^{\dagger}$, respectively. I simply write $S$ for
either $S_{A}$ or $S_{B}$. I can also
write\cite{Celeghini:1992yv,11.}:
\be\lab{M3}
  |0 (\theta) \rangle = \sum_{n=0}^{+\infty} \sqrt{W_n} \left( |n \rangle
  \otimes |{n} \rangle  \right) ~,
\ee
\be\label{M4}
  W_n = \prod_k
  \frac{\sinh^{2n_k}\theta_{k}}{\cosh^{2(n_k+1)}\theta_{k}}\,,
\ee
with $n$ denoting the set $\{ {n}_{\kappa} \}$ and with $ 0 < W_n
< 1$
 and $\sum_{n=0}^{+\infty} W_n = 1$. Then
\be\lab{M5}  \langle 0(\theta) |S_{A}|0(\theta) \rangle =
\sum_{n=0}^{+\infty} W_n ln W_n ~, \ee
and thus $S$ can be interpreted as the entropy operator
\cite{Celeghini:1992yv,11.} and it provides a measure of the
degree of entanglement. I remark that the entanglement is truly
realized in the infinite volume limit where
\be\lab{ort2} \langle 0 (\theta)| 0  \rangle = e^{-
\frac{V}{(2\pi)^{3}}\int d^{3} {\kappa} \ln \cosh \theta_{\kappa}}
 \mapbelow{V \rightarrow \infty} 0 ~, \ee
provided $\int d^{3} {\kappa} \ln \cosh \theta_{\kappa}$ is  not
identically zero. The probability of having the component state
$|n \rangle
  \otimes |{n} \rangle$  in the state $|0 (\theta) \rangle$ is
$W_n$. Since $W_n$ is a decreasing monotonic function of $n$, the
contribution of the states $|n \rangle
  \otimes |{n} \rangle$ would be
suppressed for large $n$ at finite volume. In that case, the
transformation induced by the unitary operator $G^{-1}(\theta)
\equiv \exp(- i \sum_{\kappa}\theta_{\kappa}
 {\cal G_{\kappa}})$ could disentangle the $A$ and $B$
sectors. However, this is not the case in the infinite volume
limit, where the summation extends to an infinite number of
components and Eq. (\ref{ort2}) holds (in such a limit Eq.
(\ref{19}) is only a formal relation since $G^{-1}(\theta)$ does
not exist as a unitary operator).

It is interesting to note that, although the mode $B$ is related
with quantum noise effects, nevertheless the $A-B$ (or
system-environment) entanglement is not affected by such noise
effects. The robustness of the entanglement is rooted in the fact
that, once the infinite volume limit is reached, there is no
unitary generator able to disentangle the system-environment
coupling.

\section{Chaotic behavior of the trajectories in the $\cal H$ space}

In order to discuss the chaotic behavior under certain conditions
of the trajectories in the $\cal H$ space, it is useful to recall
some of the features of the $SU(1,1)$ group structure. See, e.g.,
\cite{perelomov} for more details.

$SU(1,1)$ realized on ${\it C} \times {\it C}$ consists of all
unimodular $2 \times 2$ matrices leaving invariant the Hermitian
form $|z_{1}|^{2} - |z_{2}|^{2}$, $z_{i} \in {\it C}, i=1,2$. The
complex $z$ plane is foliated under the group action into three
orbits: $X_{+} = \{z:|z|<1 \}$, $X_{-} = \{z:|z|>1 \}$ and $X_{0}
= \{z:|z|=1 \}$.

It may be shown that the unit circle $X_{+} = \{\zeta : |\zeta|<1
\}$, $\zeta \equiv e^{i \phi} \tanh \theta$, is isomorphic to the
upper sheet of the hyperboloid which is the set $\bf H$ of
pseudo-Euclidean bounded (unit norm) vectors ${\bf n}: {\bf n}
\cdot {\bf n} = 1$. $\bf H$ is a K\"ahlerian manifold with metrics
$ds^{2} = 4\frac{{\partial}^{2} F}{\partial \zeta
\partial {\bar{\zeta}}} d\zeta \cdot d \bar{\zeta}$. Here $F
\equiv -\ln(1 - {|\zeta|}^{2})$ is the K\"ahlerian potential and
the metric is invariant under the group action.

The K\"ahlerian manifold $\bf H$ is known to have a symplectic
structure and thus it may be considered as the phase space for the
classical dynamics generated by the group action\cite{perelomov}.

The $SU(1,1)$ generalized coherent states are recognized to be
``points" in $\bf H$ and transitions among these points induced by
the group action are therefore classical
trajectories\cite{perelomov} in $\bf H$ (a similar situation
occurs\cite{perelomov} in the $SU(2)$ (fermion) case).

The above considerations thus show that the space of the unitarily
inequivalent representations of the ccr is a K\"ahlerian manifold,
${\cal H} \equiv \{ {\cal H}_\theta, ~ \forall \theta \} \approx
{\bf H}$, namely it has a symplectic structure and a classical
dynamics is established on it by the $SU(1,1)$ action (generated
by $\cal G$ or by $p_{\theta}$ in the notation of Section 1:
${\cal H}_{\theta} \rightarrow {\cal H}_{\theta'}$): Trajectories
in ${\cal H}$ describe transitions through the representations
${\cal H}_{\theta} = \{{|0 (\theta)\rangle}\}$ as the
$\theta$--parameter changes (i.e. through the physical phases of
the system, the system order parameter being dependent on
$\theta$). One may then assume time-dependent $\theta$: $\theta =
\theta (t)$. For example, this is the case of dissipative systems
and of non-equilibrium thermal field theories where
$\theta_{\kappa} = \theta_{\kappa} [\beta(t)]$, with $\beta$ the
inverse time-dependent temperature.

In conclusion, the group action induces classical trajectories in
$\cal H$. Such a result has been also obtained
elsewhere\cite{Manka,DelGiudice} on the ground of more
phenomenological considerations.

In the following it will be convenient to use the notation $|0
(t)\rangle_{\theta} \equiv |0 (\theta (t))\rangle$. For any
$\theta(t) = \{\theta_{\kappa}(t), \forall \kappa \}$ it is
\be\lab{nr} {}_{\theta}\langle 0(t) | 0(t)\rangle_{\theta}  = 1
~,~~\forall t~. \ee
I will now restrict the discussion to the case in which
$\theta_{\kappa} (t)$ is, for any $\kappa$, a growing function of
time and
\be\lab{cond} \theta (t) \neq \theta (t')~, ~~\forall  t \neq
t',~~~ {\rm and} ~~~\theta (t) \neq \theta^{\prime} (t')
~,~~\forall t,t' ~.
\ee
Under such conditions, as shown below, the trajectories in $\cal
H$ satisfy the requirements for chaotic behavior in classical
nonlinear dynamics. These requirements can be formulated as
follows \cite{hilborn}:

i)~~ the trajectories are bounded and each trajectory does not
intersect itself.

ii)~~trajectories specified by different initial conditions do not
intersect.

iii) trajectories of different initial conditions are diverging
trajectories.

Let the initial time be $t_{0} = 0$. The trajectory "initial
condition" is then specified by the $\theta(0)$-set, $\theta(0) =
\{\theta_{\kappa}(0), \forall \kappa \}$. One obtains
\be\lab{tt} {}_{\theta}\langle 0(t) | 0(t') \rangle_{\theta}
\mapbelow{V \rightarrow \infty} 0 ~, ~~ \forall \, t\, , t' ~ ,
~~~ {\rm with}~~~t \neq t' ~ , \ee
provided $ {\int \! d^{3} \kappa \, \ln \cosh
(\theta_{\kappa}(t)-\theta_{\kappa}(t'))}$ is finite and positive
for any $t \neq t'$ . Eq. (\ref{tt}) expresses the unitary
inequivalence of the states $|0(t)\rangle_{\theta} $ (and of the
associated Hilbert spaces $\{| 0(t) \rangle_{\theta} \}$) at
different time values $t \neq t'$ in the infinite volume limit.
The non-unitarity of time evolution implied for example by the
damping is consistently recovered in the unitary inequivalence
among representations $\{| 0(t) \rangle_{\theta} \}$'s at
different $t$'s in the infinite volume limit.

The trajectories are  bounded in the sense of Eq. (\ref{nr}),
which shows that the ``length" (the norm) of the ``position
vectors" (the state vectors at time $t$) in $\cal H$ is finite
(and equal to one) for each $t$. Eq. (\ref{nr}) rests on the
invariance of the Hermitian form $|z_{1}|^{2} - |z_{2}|^{2}$,
$z_{i} \in {\it C}, i=1,2$ and I also recall that the manifold of
points representing the coherent states $|0(t)\rangle_{\theta}$
for any $t$ is isomorphic to the product of circles of radius
${r_{\kappa}}^{2} = \tanh^{2}(\theta_{\kappa}(t))$ for any
$\kappa$ , as mentioned above.

I also observe that Eq. (\ref{tt}) expresses the fact that the
trajectory does not crosses itself as time evolves (it is not a
periodic trajectory): the ``points" $\{|0(t)\rangle_{\theta}\}$
and $\{|0(t')\rangle_{\theta}\}$ through which the trajectory
goes, for any $t$ and $t'$, with $t \neq t'$, after the initial
time $t_{0} = 0$, never coincide. The requirement $i)$ is thus
satisfied.

In the infinite volume limit, we also have
\be\lab{ttn} {}_{\theta}\langle 0(t) | 0(t') \rangle_{\theta ' }
\mapbelow{V \rightarrow \infty} 0 \quad ~~ \forall \, t\, , t'
\quad , ~~\forall \, {\theta} \neq {\theta^{\prime}} ~. \ee
Notice that, under the assumption (\ref{cond}), Eq. (\ref{ttn}) is
true also for $t = t'$. The meaning of Eqs. (\ref{ttn}) is that
trajectories specified by different initial conditions
${\theta}(0) \neq {\theta^{\prime}(0)}$ never cross each other.
The requirement ii) is thus satisfied.

Let me now study how the ``distance" between trajectories in the
space $\cal H$ behaves as time evolves. I consider two
trajectories of slightly different initial conditions, say
${{\theta}^{\prime} (0)} = {\theta} (0) + \delta \theta$, with
small $\delta \theta$. A difference between the states $|
0(t)\rangle_{\theta}$ and $| 0(t)\rangle_{\theta^{\prime}}$ is the
one between the respective expectation values of the  number
operator $A_{\kappa}^{\dagger} A_{\kappa}$. We have, for any
$\kappa$ at any given $t$,
$$
\Delta {\cal N}_{A_{\kappa}}(t) \equiv {\cal
N'}_{A_{\kappa}}\bigl(\theta'(t)\bigr ) - {\cal
N}_{A_{\kappa}}\bigl(\theta(t)\bigr ) =
{}_{\theta^{\prime}}\langle 0(t) | A_{\kappa}^{\dagger} A_{\kappa}
|0(t)\rangle_{\theta^{\prime}} - {}_{\theta}\langle 0(t)
|A_{\kappa}^{\dagger} A_{\kappa} | 0(t) \rangle_{\theta}
$$
\be\lab{co1} = \sinh^{2}\bigl ( {\theta^{\prime}}_{\kappa}(t)
\bigr ) - \sinh^{2}\bigl ( {\theta}_{\kappa}(t) \bigr ) = \sinh
\bigl ( 2{\theta}_{\kappa}(t)\bigr ){\delta \theta_{\kappa}(t)} ~,
\ee
where ${\delta \theta_{\kappa}(t)} \equiv
{\theta^{\prime}}_{\kappa}(t) - {\theta}_{\kappa}(t)$ is assumed
to be greater than zero, and the last equality holds for ``small"
${\delta \theta_{\kappa}(t)}$ for any $\kappa$ at any given $t$.
Then, by assuming that $\frac{\partial{\delta
\theta_{\kappa}}}{\partial t}$ has negligible variations in time,
the time-derivative gives
\be\lab{co3} \frac{\partial}{\partial t} \Delta {\cal
N}_{A_{\kappa}}(t) = 2 \frac{\partial
{\theta}_{\kappa}(t)}{\partial t }\cosh \bigl ( 2
{\theta}_{\kappa}(t)  \bigr ){\delta \theta_{\kappa}} ~, \ee
which shows that, provided $\theta_{\kappa}(t)$ is a growing
function of $t$, small variations in the initial conditions lead
to growing in time $\Delta {\cal N}_{A_{\kappa}}(t)$, namely to
diverging trajectories as time evolves.

In the assumed hypothesis, at enough large $t$ the divergence is
dominated by $\exp{(2\theta_{\kappa}(t)) }$. For each $\kappa$,
the quantity $2\theta_{\kappa}(t) $  could be thus thought to play
the role similar to the one of the Lyapunov exponent.

I also observe that, since\cite{Celeghini:1992yv} $\sum_{\kappa}
E_{\kappa} {\dot {\cal N}}_{A_{\kappa}} dt = \frac{1}{\beta}
dS_{A}$, where $E_{\kappa}$ is the energy of the mode
$A_{\kappa}$,  $dS_{A}$ is the entropy variation associated to the
modes $A$ and ${\dot {\cal N}}_{A_{\kappa}}$ denotes the time
derivative of ${\cal N}_{A_{\kappa}}$,   the divergence of
trajectories of different initial conditions may be expressed in
terms of differences in the variations of the entropy (cf. Eqs.
(\ref{co1}) and (\ref{co3})):
\be\lab{co4} \Delta \sum_{\kappa} E_{\kappa}{\dot {\cal
N}}_{A_{\kappa}}(t) dt  = \frac{1}{\beta} \bigl ( dS'_{A} - dS_{A}
\bigr ) ~. \ee
In conclusion, also the requirement iii) is satisfied. The
trajectories in the $\cal H$ space thus exhibit, under the
condition (\ref{cond}), properties typical of the chaotic behavior
in classical nonlinear dynamics.

\section{Conclusions}

The deformed Hopf algebra structure of QFT implies the doubling of
the system degrees of freedom. The doubled degrees of freedom play
the role of the thermal bath or environment degrees of freedom and
are entangled with the system degrees of freedom. They also
account for quantum noise in the fluctuating random forces in the
system--environment coupling. In such a frame, the trajectories in
the space of the representations of the canonical commutation
relations turn out to be classical trajectories and, under
convenient conditions, they may exhibit properties typical of
classical chaotic trajectories in nonlinear dynamics. Finally, it
might be worth to remark that quantum noise effects accounted for
by the doubled degrees of freedom are not the source of such a
chaotic behavior.

\section*{Acknowledgements}

Partial financial support from Murst, INFN, INFM and the ESF
Program COSLAB is acknowledged.

\end{document}